# Boundary conditions for magnetization in magnetic nano-elements

K.Yu. Guslienko[1*], A.N. Slavin[2]

## Abstract

We show that the dynamic magnetization at the edges of a thin magnetic element with finite lateral size can be described by new effective boundary conditions that take into account inhomogeneous demagnetizing fields near the element edges. These fields play a dominant role in the effective pinning of the dynamic magnetization at the boundaries of mesoscopic and nano-sized magnetic elements. The derived effective boundary conditions generalize well-known Rado-Weertman boundary conditions and are reduced to them in the limiting case of a very thin magnetic element.

[1] Materials Science Division, Argonne National laboratory, Argonne, IL 60439
[*] To whom correspondence should be addressed. E-mail: Gusliyenko@anl.gov
[2] Department of Physics, Oakland University, Rochester, Michigan 48309

Rapid progress in magnetic data recording and sensor technologies creates a motivation to work with submicron magnetic elements [1]. The physics of mesoscopic and nano-magnetic elements is qualitatively different from that of bulk magnetic systems. Confinement of spin wave modes and other finite-size effects dominate the properties of magnetic nano-particles [2, 3] and create opportunities for novel applications in spintronic devices [4, 5]. The use of small magnetic elements in data recording [4, 6] or for current-induced microwave generation and switching [5, 6] depends on our understanding their fundamental dynamical properties.

The central problem is to understand the dipole-dipole interaction and its interplay with other factors, including the exchange interaction and the surface anisotropy. When the relevant interactions are properly taken into account, it is possible to calculate the excitation spectra of the magnetic elements in terms of spin wave eigenmodes. These spectra provide information on the characteristic times of magnetization reversal as studied experimentally [7, 8], and provide much needed general insights.

The magnetizations dynamics of a magnetic element can be described using the Landau-Lifshits equation of motion. This approach contains contributions from the non-uniform exchange interaction, as well as from the long-range dipole-dipole interaction, which is also non-uniform for non-ellipsoidal magnetic elements. The eigenfrequencies and eigenmode distributions of spin-wave excitations obtained depend strongly on the surface boundary conditions. Knowledge of these boundary conditions is important to calculate the linear spin wave spectra, to analyze the magnetization (**M**) reversal, and to investigate non-linear topological excitations (magnetic vortices), etc.

It is known that the usual electrodynamic boundary conditions of the Maxwell classical theory leave the amplitude of the dynamic magnetization at the boundary undefined. Maxwell's theory requires the continuity of the normal components of the magnetic induction and the tangential components of the magnetic field. The problem is that the magnetic moments near the boundaries experience the influence of local magnetic fields that are different from the fields in the bulk. Kittel [9] introduced boundary conditions of total "pinning" (**M**=0 at the boundary) based on Neel's concept of surface anisotropy [10] to explain experimental data on spin-wave resonances in magnetic films. The general "exchange" boundary conditions for the dynamic magnetization were then formulated by Rado and Weertman (RW) [11]. In addition to Kittel's term, RW took into account the influence of the exchange interaction, and obtained what is known as the Rado-Weertman boundary conditions:

$$L_e^2 \mathbf{M} \times \frac{\partial \mathbf{M}}{\partial n} + \mathbf{T}_s = 0, \qquad (1)$$

where $L_e = (2A/M_s^2)^{1/2}$ is the characteristic exchange length of a material (defining the length scale at which the exchange interaction becomes important), $A$ is the exchange stiffness, $M_s$ is the saturation magnetization, $\partial/\partial n$ is the partial derivative along a unit vector **n** (inward normal direction to the particle surface), and $\mathbf{T}_s$ is the sum of all the surface torques which arise from forces other than the exchange interaction. The term $\mathbf{T}_s$ usually contains contributions from the Neel surface anisotropy $\mathbf{T}_a$, but contributions from other local fields are also possible. The boundary conditions (1) generalize Kittel's, and permit both a totally "pinned" magnetization (**M**=0) when the torque $\mathbf{T}_s$ is large, and also a pinning of an arbitrary magnitude up to the totally "free" or "unpinned" magnetization ($\partial \mathbf{M}/\partial n = 0$) at the boundaries when $\mathbf{T}_s$ is small and the exchange interaction is dominant.

In the present article we demonstrate that to derive accurate boundary conditions for the dynamic magnetization in a mesoscopic or nano-sized magnetic element it is not sufficient to take into account only the Neel surface anisotropy in the expression for $\mathbf{T}_s$. It is also necessary to include a contribution from the strongly non-uniform internal dipolar field existing near the edges. Then, the surface torque in Eq. (1) becomes $\mathbf{T}_s = \mathbf{T}_a + \mathbf{T}_m$, where the second dipolar (or magnetostatic) term can become dominant in a certain size range. It is then convenient to rewrite Eq. (1) in the form:

$$\mathbf{M} \times \left( L_e^2 \frac{\partial \mathbf{M}}{\partial n} - \nabla_M E_a + \mathbf{H}_m L \right) = 0, \qquad (2)$$

where $E_a(\mathbf{M})$ is the energy density of the surface anisotropy, and $\mathbf{H}_m$ is the dipolar field near the edge. To explicitly illustrate the general boundary conditions (2), we consider below the case of an uniaxial surface anisotropy with anisotropy constant $K_s$ and the anisotropy axis direction given by an unit vector $\mathbf{n}_a$. The effective field of surface anisotropy is then given by $\mathbf{H}_a = -\nabla_M E_a = (2K_s / M_s^2)(\mathbf{M} \cdot \mathbf{n_a})\mathbf{n_a}$. Our main task will be to evaluate the dipolar field $\mathbf{H}_m$ that exists near the edge of a thin magnetic element.

We represent the magnetization in Eq. (2) in the form $\mathbf{M} = M_s \mathbf{i}_0 + \mathbf{m}$, where $\mathbf{i}_0$ is an unit vector in the direction of the equilibrium magnetization $\mathbf{M}_s$. We assume that the dynamic magnetization $m$ is much smaller than the $M_s$ ($m \ll M_s$) and that $\mathbf{m}$ is perpendicular to $\mathbf{M}_s$ or $\mathbf{m} \cdot \mathbf{i}_0 = 0$. In a thin in-plane magnetized element made of a soft magnetic material, vector $\mathbf{i}_0$ lies in the plane of the element, and is directed along its lateral edge to minimize the magnetostatic energy of the static magnetic configuration. The element could have an arbitrary shape. The only critical requirement is that it is thin, *i.e.* that the element thickness $L$ is of the order of the exchange length $L_e$, and is much smaller than the element lateral size. Typical shapes could be a thin rectangular prism or a thin circular or elliptical cylinder. Sub-micron plane magnetic dots prepared by lithographic patterning of thin magnetic films made of Fe, Co, or NiFe are good examples of such thin magnetic elements [8, 12, 13].

In practical calculations it is usually assumed that the dynamic magnetization at the edges is totally pinned as is described by Kittel's boundary conditions (see *e.g.* Ref. 14). But this approach is rather arbitrary, and is not based on an exact knowledge of the behavior of the dynamic magnetization near the boundary. To derive boundary conditions for a thin magnetic element we evaluate the inhomogeneous dipolar field $\mathbf{H}_m$ directly from Maxwell's equations. For mathematical simplicity we consider a case of an axially magnetized stripe with a rectangular cross-section (see Fig.1) which has only *one finite lateral caliper* (width *w*). The general solution of Maxwell's equations for the dipolar field $\mathbf{H}_m$ can be written as a sum of two fields, resulting from surface and volume magnetic charges [15]:

$$\mathbf{H}_m(\mathbf{r}) = -\nabla_\mathbf{r} \int dS \frac{m_n(\mathbf{r'})}{|\mathbf{r}-\mathbf{r'}|} + \nabla_\mathbf{r} \int dV \frac{\nabla_{\mathbf{r'}} \cdot \mathbf{m}(\mathbf{r'})}{|\mathbf{r}-\mathbf{r'}|} \quad . \tag{3}$$

We evaluate below the dipolar field (3) for an infinitely long magnetic stripe having thickness *L*, uniformly magnetized along the *y*-axis, while the *z*-axis is directed along the stripe thickness (see Fig. 1). The boundary conditions for magnetization in a stripe can be written as projections of the vector torque equation (2) on the coordinate axes. Since the stripe is infinite in the *y* direction the distribution of the dynamic magnetization $\mathbf{m}(\mathbf{r}) = \mathbf{m}(x, z)$ within the stripe and the demagnetizing field $\mathbf{H}_m(\mathbf{r}) = \mathbf{H}_m(x, z)$ are independent of the coordinate *y*, and *y*-components of both these vectors are vanishing. We can also assume a homogeneous distribution of the dynamic magnetization along the coordinate *z* (making $\mathbf{m}(\mathbf{r}) = \mathbf{m}(x)$), since we are considering thin magnetic elements with the thickness of the order of $L_e$. For the case of a thin magnetic stripe with aspect ratio $p = L/w \ll 1$ and situation when $\mathbf{m}(\mathbf{r}) = \mathbf{m}(x)$, the torque (2) and the dipolar field (3) have only *x* and *z* components, and can be simplified. At first, we consider only the *x*-component $H_{mx}(\mathbf{r})$ of the dipolar field (3). Since we are interested in the boundary conditions at the lateral edges of the stripe (the planes $x = \pm w/2$ in Fig. 1), we can also average the *x*-component of the dipolar field (3) over the coordinates *y* and *z*, making it a function of the coordinate *x* only: $h(x) = \langle H_{mx}(\mathbf{r}) \rangle_{y,z}$. We separate the contributions from the surface and volume magnetic charges, and write $h(x) = h_S(x) + h_V(x)$. Note, that at the lateral surface $x = w/2$ the surface charges are given by $m_x(w/2)$ and the volume charges are $\partial m_x(x)/\partial x$.

Evaluation of the integrand in the first (surface) integral in (3) at the face surfaces ($z=0, L$) of the stripe shows that the face surface magnetic charges (proportional to $m_z$) do not contribute to the $x$-component of the dipolar field $h(x)$ [16]. Evaluation of the same integrand at the lateral ($x = \pm w/2$) surfaces of the stripe yields the dipolar field in the form $h_S(x) = [2\pi\theta(x - w/2) + F((w/2 - x)/L)]m_x(w/2)$, where $F(\eta) = 2\eta \ln(1 + 1/\eta^2) + 4\tan^{-1}\eta$. Direct calculation shows that near the lateral boundary $x = w/2$ of the stripe the contribution of the second term in the expression for $h_S(x)$ is small (of order of the stripe aspect ratio $p$). Only the first term of $h_S(x)$ gives contribution (of order of $2\pi$) to the boundary conditions in the main approximation. In the calculation of the dipolar field $h_S(x)$ near the right $x \approx w/2$ lateral surface of the stripe we also neglect the contribution to it from the surface charges at the left lateral surface $x = -w/2$ and vice versa.

To evaluate the second (volume) integral in (3) near the lateral boundary of the stripe we expand the variable magnetization $\mathbf{m}(x')$ in a Taylor series. Direct calculation shows that only the $x$ component of the variable magnetization is important in this case, and that the main contribution to the dipolar field $h_V(x)$ comes from term containing the first spatial derivative of $m_x$. This yields the "volume" dipolar field in the form $h_V(x) = wI(x, p)\partial m_x(x)/\partial x$, where near the boundary $x=w/2$ the function $I(x, p)$ is evaluated as $I(w/2, p) = p - 2p\ln p$ [16]. All others terms in the field $h_V(x)$, containing higher order derivatives $\partial^s m_x(x)/\partial x^s$ ($s>1$), in the limit of a thin stripe $p << 1$ are substantially smaller than the first term [16]. A similar situation exists at the other lateral boundary $x = -w/2$. Due to the strong non-uniformity of the dipolar field near the lateral surfaces of the stripe, the integrals of the dipolar field near the lateral surfaces of the stripe can be evaluated as $\int h(x)dx = h(w/2)L$.

Substituting the calculated expressions of $h(x) = h_S(x) + h_V(x)$ for $\mathbf{H}_m$ in Eq. (2), we get the following relations between the x-component of $\mathbf{m}$ and the first derivative of this component at the stripe boundaries:

$$\frac{\partial m_x(\xi)}{\partial \xi} \pm d(p,L) m_x(\xi)\Big|_{\xi=\pm 1/2} = 0 . \qquad (4)$$

These relations that can be interpreted as effective boundary conditions, where $d(p, L)$ is the effective "pinning" parameter and $\xi = x/w$ is the dimensionless coordinate perpendicular to the lateral boundary of the stripe. The direction of the normal $n$ to the lateral surface of the stripe is defined as $\mathbf{n} = -x\mathbf{e}_x$, where $\mathbf{e}_x$ is unit vector along the x-axis. Similar boundary conditions can be obtained for the z-component of the dynamic magnetization.

As an alternative example of a thin magnetic element we also considered a circular cylinder having thickness $L$ and radius $R$ ($L<<R$). We used an approach similar to that for long rectangular stripe and obtained a result for the pinning parameter analogous to the result obtained for the stripe. A general expression for the pinning parameter in (4), which is correct in both rectangular and cylindrical geometry, can be written in the following form:

$$d(p,L) = \frac{2\pi\left[1 - \left(\frac{K_S}{\pi M_S^2 L}\right)\right]}{p\left[a + b\ln(1/p) + \left(\frac{L_e}{L}\right)^2\right]}, \qquad (5)$$

where the values of the coefficients $a$ and $b$ are determined by the geometry of the element ($a, b \sim 1$), and the parameter $p$ is the thickness to lateral size aspect ratio of a thin magnetic element. In particular, $p=L/w$ for a stripe of the width $w$ and $p = L/R$ for a circular cylinder of the radius $R$. Calculations yield the following values for coefficients $a$ and $b$: $a=1$, $b=2$ for a rectangular stripe, and $a = 2(6\ln 2 - 1)$, $b=4$ for a circular cylinder.

The above derived boundary conditions (4) with the effective pinning parameter (5) generalize the well-known RW boundary conditions [11] for the case of a thin magnetic element having a finite lateral size, and represent the main result of this paper. The sign of the pinning parameter in our definition (5) is *opposite* to the sign in the pinning parameter defined in [11], so that $d(p,L) > 0$ in (5) corresponds to the "easy plane" type of *effective* surface anisotropy. The calculated pinning parameter (5) corresponds to a strong pinning if $w, R >> L \geq L_e$ (dipolar dominated regime), and to a weak pinning if $(Lw)^{1/2} < L_e$ or $(LR)^{1/2} < L_e$ (exchange dominated regime). We believe that although our calculations were done for rectangular and circular geometries, similar effective boundary conditions could be obtained in other geometries, in particular for a thin magnetic dot having the shape of an elliptical cylinder.

The pining parameter $d(p,L)$ calculated from Eq. (5) is shown in Fig. 2 as a function of the element thickness. The exchange interaction dominates for small $L$, and the pinning parameter decreases as the element thickness $L$ decreases. In the case of a non-zero surface anisotropy, deviations of the pinning parameter (5) from the purely dipolar pinning [16] occur for the element thickness $L < 10$ nm and are stronger for an "easy axis" type of surface anisotropy ($K_s>0$) (see dot-dashed line in Fig. 2). For this type of surface anisotropy the pinning parameter (5) also strongly differs from the RW pinning due to the competing contributions from surface anisotropy and dipolar interaction. For $K_s=0$ the pinning (5) depends on the ratio $L/L_e$ and is strong if $L/L_e > 0.1$. For lager values of $L$, but still in the limit $p << 1$, the dipolar contribution to pinning becomes dominant independently of the sign and value of the surface anisotropy. The absolute value of $|K_s|=0.20$ erg/cm$^2$ used in Fig. 2 ($M_s$= 800 G) corresponds to a relatively strong surface anisotropy. We note that for the majority of soft magnetic materials the contribution from the surface anisotropy to the effective pinning parameter (5) can be neglected as usually we have $K_s << \pi L M_s^2$.

Fig. 3 demonstrates crossing regime from the case of a strong dipolar pinning to the case of a weaker exchange-dominated pinning when the element thickness is decreased and the element aspect ratio $p$ is kept constant (proportional scaling of the element's sizes). The pinning vanishes at $L \to 0$, which reflects the increasing role of the short-range exchange interaction. Note, that purely dipolar value of the pinning parameter $d(p) = 2\pi p^{-1}[1 + 2\ln(1/p)]$ derived in Ref. 16 and RW value of the pinning $d_{RW}(p,L) = -2K_s L / p M_s^2 L_e^2$ [11], serve as two asymptotes for the pinning parameter given by Eq. (5). The first limiting value is achieved for $L > L_e$, while the second limiting value we get for $L < L_e$. It is also clear, that the general pinning parameter (5), in the limit $L \to 0$, is equivalent to the classic RW pinning, independently of the sign of the surface anisotropy.

We stress, that, although the boundary conditions (4) look formally analogous to the exchange boundary conditions in a perpendicularly magnetized film, the effective pinning is actually a result of the interplay of the exchange, dipolar, and surface anisotropy terms. In contrast to the usual "exchange" pinning, the pinning parameter (5) is not fully determined by the surface anisotropy of the magnetic material. The physical role of this generalized pinning is to minimize the total surface energy (sum of exchange, anisotropy and magnetostatic energies). The magnetostatic part results from the induced surface charges $\sigma = (\mathbf{m} \cdot \mathbf{n})_S$ at the edges of a finite-size non-ellipsoidal magnetic element and volume charges near its edges. For $p \to 0$ the pinning parameter $d(p, L)$ of Eq. (5) is rather large, and the boundary conditions (4) are close to the Kittel's boundary conditions [9] that were traditionally used at the lateral edges of thin magnetic elements [14]. A more detailed analysis shows that in the boundary conditions (4) the term proportional to the dynamic magnetization **m** comes from the surface magnetic charges and surface anisotropy. The term proportional to the derivative $\partial m/\partial x$ comes from the volume magnetic charges (and exchange), if we retain only the main terms in the small parameter $p$. The strong pinning in the dipole-dominated regime corresponds to a large ratio of surface/volume charges, and for a thin magnetic elements ($p \ll 1$) represents a "finite-size" effect.

The derived boundary conditions (4) are especially important within the thickness range 2-10 nm (see Fig. 2, 3), where the pinning described by Eq. (5) differs significantly from the "dipolar" value given by horizontal line in Fig. 3 (see also [16]). Our predictions can be tested in any dynamical experiments using magnetic elements with the aspect ratio $p \ll 1$ and $L \leq 10$ nm. In particular, for the conditions of the experiment [7], where the "free" layer of a nano-pillar driven by spin-polarized current has a shape of a thin elliptical cylinder with the sizes 3×130×70 nm, our equation (5) predicts negligible pinning at the lateral edge of the "free" layer, thus excluding the influence of the exchange interaction on the frequency of the lowest spin wave mode excited in the nano-pillar. This conclusion is supported by the results of the experiment [7], where the dependence of the current-induced microwave frequency on the bias magnetic field for the small precession angle regime is well described by the purely dipolar (non-exchange) expression for the quasi-homogeneous precession mode of a nano-pillar. We also note that in the magnetic elements of the thickness of $L = 30$ nm and the in-plane size of about $w = 1000$ nm Eq. (5) predicts strong "dipolar" pinning for the lowest spin wave modes, and this result is also well-supported by experiments (see *e.g*. Fig. 4 in [16]).

In summary, we derived general boundary conditions (4) with the *effective* pinning parameter (5) for the dynamic magnetization of thin magnetic non-ellipsoidal elements. These conditions take into account exchange interaction, surface anisotropy, and non-uniform dipolar field near the element lateral edges. We have shown that in the range of the element thickness $L_e < L \ll w, R$ the contribution of the dipolar field to the effective pinning parameter is dominant, while for $L \to 0$ ($L_e > L > 0$) the exchange and surface anisotropy contributions become gradually more important, and our boundary conditions are reduced to the well-known Rado-Weertman form [11]. The derived general boundary conditions (4), (5) are important in the interpretation of spin wave spectra of nano-sized magnetic elements, and are well-supported by several independent experiments (see *e.g.* [7] and [16]).

We would like to thank S.D. Bader for his useful and constructive comments on the manuscript. The U.S. Department of Energy, BES Material Sciences under Contract No. W-31-109-ENG-38 supported the work at Argonne National Laboratory. The work at Oakland University was supported by the Grant # W911NF-04-1-0299 from the U.S. Army Research Office, by the MURI grant W-911NF-04-1-0247 from the Department of Defense, and by the Oakland University Foundation.

**Captions for illustrations**

for the manuscript: "Boundary conditions for magnetization in magnetic nano-elements" by K.Yu. Guslienko, and A.N. Slavin

Fig. 1. Coordinates system for a thin magnetic stripe.

Fig. 2. Pinning parameter for thin magnetic stripe vs. element thickness $L$ (the stripe width is constant $w$=1000 nm). The dashed line corresponds to pure dipolar pinning ($K_s$ =0, $L_e$=0) of Ref. 16. The solid line corresponds to the "easy-plane" surface anisotropy ($K_s/\pi M_s^2 = -1$), while the dot-dashed line corresponds to the "easy-axis" type of surface anisotropy ($K_s/\pi M_s^2 = 1$ nm). The dotted line corresponds to $K_s$=0, the horizontal lines correspond to RW pinning with $K_s/\pi M_s^2 = -1$ nm (upper) and $K_s/\pi M_s^2 = 1$ nm (lower). $L_e$ = 20 nm.

Fig. 3. Pinning parameter for thin magnetic stripe vs. element thickness $L$ keeping the aspect ratio $p$=0.01 as constant. The dotted line corresponds to pure dipolar pinning ($K_s$ =0, $L_e$ =0). The solid line and the dashed line are for a surface anisotropy $K_s/\pi M_s^2 = -2$ nm and $K_s$=0, respectively. The dashed-dotted line corresponds to RW pinning with the same parameters. $L_e$ = 20 nm.

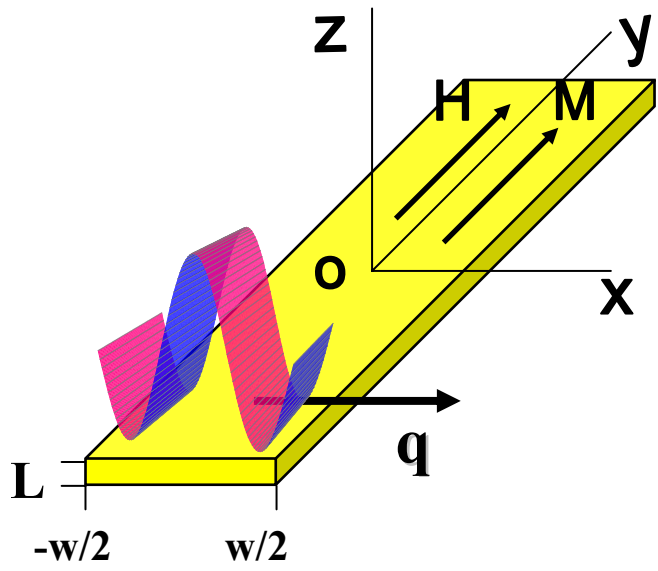

Fig. 1.

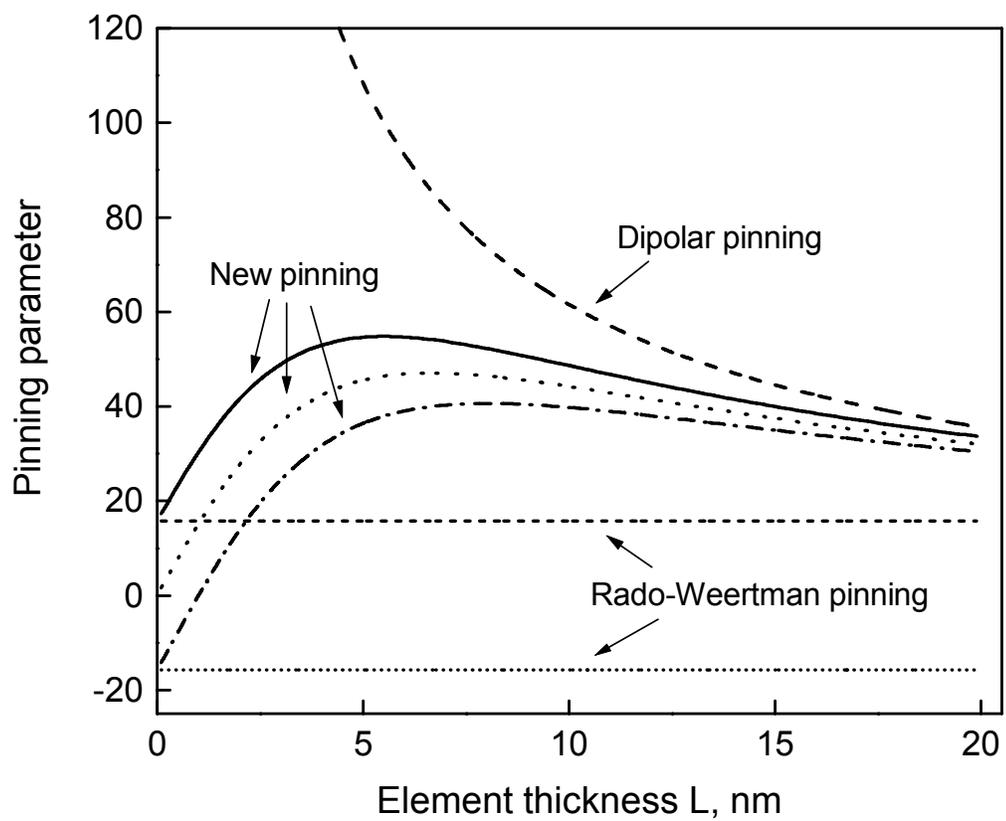

Fig. 2.

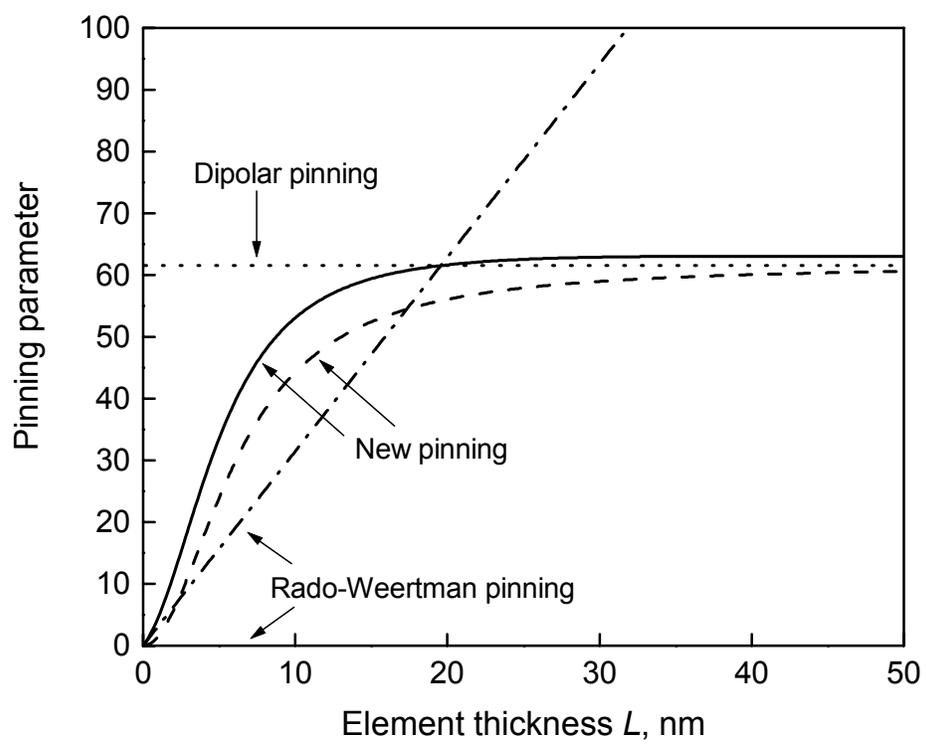

Fig. 3.